\documentclass[apj,iop]{emulateapj}

\usepackage{natbib}
\usepackage{longtable}
\usepackage{hyperref}
\usepackage{xcolor}

\shorttitle{Structures of Dwarf Galaxies with AGNs}
\shortauthors{Kimbrell et al.}

\begin{document}

\title{A Comparison Between the Morphologies and Structures of Dwarf Galaxies with and without Active Massive Black Holes}

\author{Seth J. Kimbrell}
\affil{eXtreme Gravity Institute, Department of Physics, Montana State University, MT 59715, USA}
\email{seth.kimbrell@montana.edu}

\author{Amy E. Reines}
\affil{eXtreme Gravity Institute, Department of Physics, Montana State University, MT 59715, USA}

\author{Jenny E. Greene}
\affil{Department of Astrophysical Sciences, Princeton University, Princeton, NJ 08544, USA}

\and

\author{Marla Geha}
\affil{Department of Astronomy, Yale University, New Haven, CT 06520, USA}

\begin{abstract}
We study the morphologies and structures of 57 dwarf galaxies that are representative of the general population of dwarf galaxies, and compare their demographics to a sample of dwarf galaxies hosting optically-selected AGNs. The two samples span the same galaxy stellar mass ($10^9 \lesssim M_\star/M_\odot \lesssim 10^{9.5}$) and color range, and the observations are well-matched in physical resolution. The fractions of irregular galaxies (14\%) and early-types/ellipticals ($\sim 18\%$) are nearly identical among the two samples. However, among galaxies with disks (the majority of each sample), the AGN hosts almost always have a detectable (pseudo)bulge, while a large fraction of the non-AGN hosts are pure disk galaxies with no detectable (pseudo)bulge. Central point sources of light consistent with nuclear star clusters are detected in many of the non-AGN hosts. In contrast, central point sources detected in the AGN hosts are on average more than two orders of magnitude more luminous, suggesting the point sources in these objects are dominated by AGN light. The preference for (pseudo)bulges in dwarf AGN hosts may inform searches for massive black holes in dwarf galaxies and attempts to constrain the black hole occupation fraction, which in turn has implications for our understanding of black hole seeding mechanisms.

\end{abstract}

\section{Introduction}
It is known that supermassive black holes (BHs) with masses up to $\sim 10^{10}$ M$_\odot$ reside in the center of massive galaxies \citep{kormendy1995,kormendy}. This includes Saggitarius A*, the BH that lives in the center of the Milky Way, with a mass of 4 $\times$ $10^6$ M$_\odot$ \citep{ghez2008}. What is not nearly as well understood is the mechanism that led to the formation of the initial seeds of supermassive BHs \citep{greene2020,volonteri,inayoshi2020,volonteri2021}. 

We know that BH seeds must have formed at very early cosmic times, since quasars are observed at redshifts as high as 7.5 \citep{Mortlock,vito, banados2018}. This indicates that BH seeds existed and grew to enormous masses within the first Gyr after the Big Bang. While several distinct mechanisms have been advanced as possible formation channels \citep[e.g.,][]{loeb1994,madau2001,begelman1978}, we still do not know which, if any, reflect what happened in the earlier Universe.

While current technology does not allow us to observe and directly study seed BHs at high redshifts, dwarf galaxies in the local Universe that host the smallest supermassive (or just ``massive") BHs give us an opportunity to study BHs that have not grown much compared to their more massive counterparts in giant galaxies. Studies regarding the demographics of BHs in dwarf galaxies (e.g., the BH occupation fraction) also facilitate constraints on the formation mechanisms of seed BHs \citep{volonteri,habouzit2016,Greene,miller2015,she2017chandra}. Dwarfs come in a wide variety of shapes and sizes, with structures ranging from very irregular galaxies to ellipticals and late-type spirals \citep{mcconnachie,kormendy2015dwarf}, and therefore understanding whether BHs appear preferentially in a particular type (or types) of dwarf galaxies is crucial as attempts are made to estimate the occupation fraction of BHs in dwarf galaxies \citep{reines2022}.

 To gain insight into what types of dwarf galaxies can host massive BHs, we previously studied the morphologies and structures of 41 dwarf galaxies ($M_\star \leq 3 \times 10^9 M_\odot$) hosting optically-selected active galactic nuclei (AGNs) using {\it Hubble Space Telescope (HST)} near-infrared imaging \citet{kimbrell2021}. Specifically, these objects were selected as AGN hosts using optical emission-line diagnostic diagrams (e.g., the BPT diagram; 
\citealt{baldwinetal1981}). In this work, we refer to these as ``optically selected," but we differentiate these from galaxies selected as AGN hosts via optical variability searches (see, e.g., \citealt{baldassare2020,burke2023}).
Of these AGN-hosting dwarf galaxies, 85\% have regular morphologies. Most of these regular galaxies are disk-dominated with small pseudobulges, although there are also a handful of bulge-like/elliptical galaxies. The remaining 15\% of the sample are irregulars, including Magellanic-types and dwarf galaxies showing signs of interactions/mergers. Ideally, we would like to probe dwarf galaxies hosting both inactive and active BHs, however the vast majority of massive BHs known in dwarf galaxies come from samples of AGNs \citep[for a review, see][]{reines2022}. 

In this work, we study the morphologies and structures of a sample of non-AGN-hosting dwarf galaxies for a comparison to the AGN sample presented in \citet{kimbrell2021}. We aim to determine if/how AGN-hosting dwarf galaxies differ from the dwarf galaxy population in general, and place AGN-hosting dwarf galaxies in the broader context of galaxy and BH evolution. We describe our sample in Section \ref{sec:sample}. We present our analysis and results in Sections \ref{sec:analysis} and \ref{sec:results}, respectively. A summary of our conclusions is in Section \ref{sec:conclusion}.

\section{Observations and Sample Selection}\label{sec:sample}

We utilize ground-based near-infrared observations from the UKIRT Infrared Deep Sky Survey (UKIDSS) Large Area Survey \citep{Lawrence} for our sample of non-AGN-hosting dwarf galaxies. UKIDSS utilizes the United Kingdom InfraRed Telescope (UKIRT), a 3.8m infrared telescope located at Mauna Kea in Hawai'i. The survey provides images in the $Z$, $Y$, $J$, $H$, and $K$ broadband filters. We use the $Y$-band images with a central wavelength of 1.03 $\mu$m for comparison to \citet{kimbrell2021}, as this closely matches the {\it HST}/WFC3 F110W IR filter with an effective wavelength of 1.15 $\mu$m used in that work. The depth of the UKIDSS data is well-matched to our {\it HST} observations with a surface brightness sensitivy of $\mu_Y \sim 22.7$ mag arcsec$^{-2}$ \citep{kelvin2012}. 

We construct our control sample of nearby dwarf galaxies that do not host optically-selected AGNs using version 0.1.2 of the the NASA Sloan Atlas (NSA), which is a catalog built from Sloan Digital Sky Survey (SDSS) observations \citep{york2000,blanton2011}. We first impose a distance cut of $\leq$ 20 Mpc, corresponding to a redshift of $\sim 0.005$. By imposing this distance cut, we ensure that the ground-based images (with an angular resolution of $\sim 1$\arcsec) have a linear resolution comparable to the {\it HST} observations \citep{kimbrell2021} of the AGN-hosting dwarf galaxies that have a median distance of 111 Mpc (linear resolution of $\sim 100$ pc in both cases). Imposing this requirement on distance left us with 2,404 objects. Stellar mass cuts were also placed such that $8.9 \leq {\rm log(M_*/M_\odot)} \leq 9.5$, so that we explore a mass range comparable to the dwarf galaxies hosting AGNs in \citet{kimbrell2021}. This leaves 229 objects. 
\begin{figure}
\includegraphics[width=\linewidth]{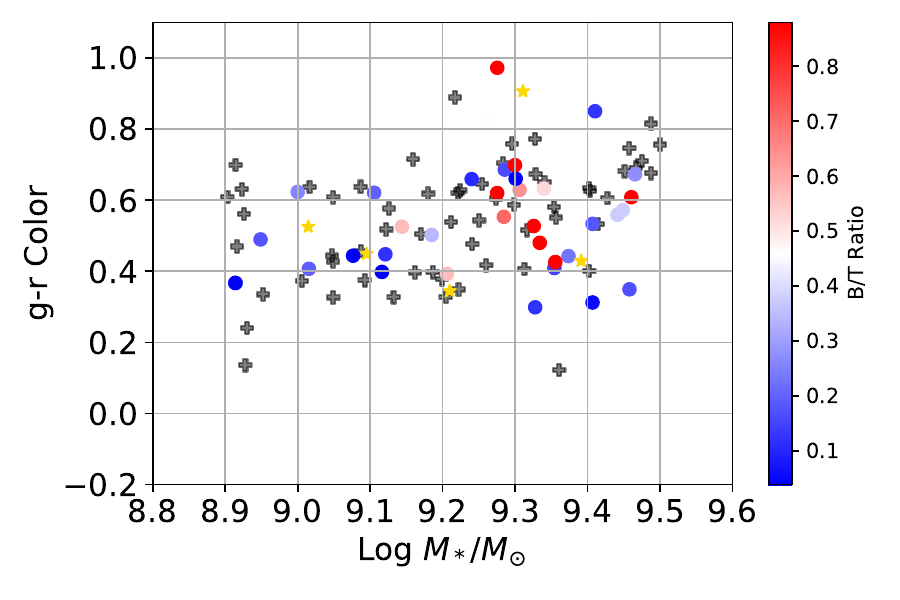}
\caption{Color-mass diagram for our sample of dwarf galaxies analyzed in this work that do not host an AGN (grey crosses) and those analyzed in \citet{kimbrell2021} that do host an AGN, demonstrating that both samples span a similar mass and color range. The AGN sample of \citet{kimbrell2021} is color coded by bulge-to-total (B/T) light ratio. Yellow stars indicate the irregular galaxies in the AGN sample.} 
\label{fig:grcolormass}
\end{figure}

\begin{figure}
\includegraphics[width=\linewidth]{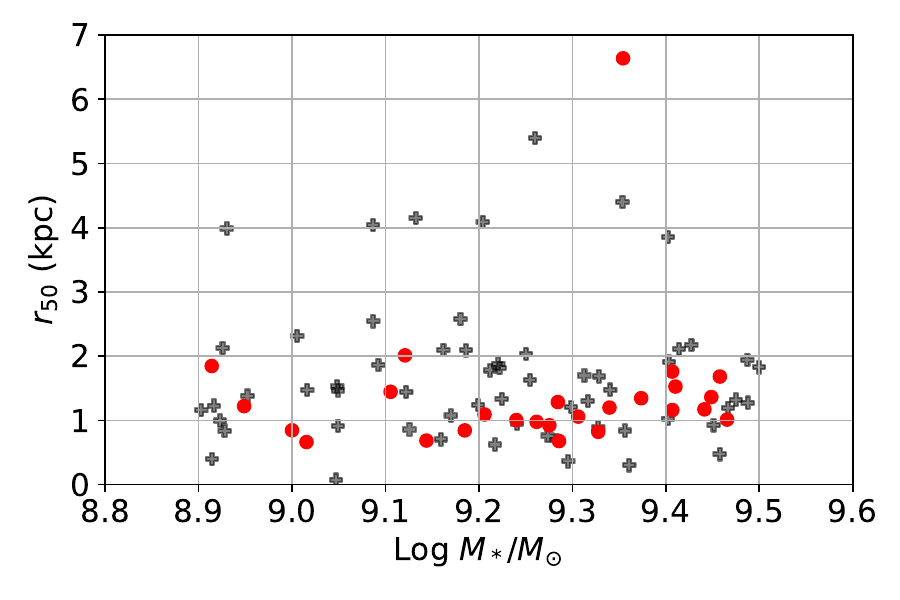}
\caption{ Size-mass diagram for our sample of dwarf galaxies analyzed in this work that do not host an AGN (grey crosses) and those analyzed in \citet{kimbrell2021} that do host an AGN (red dots), demonstrating that both samples span a similar mass and size range.}
\label{fig:sizemass}
\end{figure}

Of the 229 objects selected in the NSA, 82 had images in the UKIDSS Large Area Survey. However, upon inspection of the data, 22 galaxies did not have sufficient image quality for our analysis. Often this was because the galaxy was on the edge of the image. In other cases, there were large numbers of bad pixels in the image, or the galaxy had too low S/N to accurately analyze. In the end, we were left with 60 dwarf galaxies with high-quality images in UKIDSS. 

To minimize the chance of including AGN-hosting galaxies in this sample, we checked a variety of resources. First, none of these galaxies were selected as BPT AGNs by \citet{reines}, which used the same parent sample of dwarfs. However, two of these 60 were selected as BPT composites by \citet{reines} and so were excluded from this work. We also searched the catalog of 237 low-mass variability-selected AGNs presented in \citet{baldassare2020} and found none of the galaxies analyzed in this work. Finally, 18 of the galaxies in our sample are covered by Chandra observations; one of these dwarfs has an X-ray detection in the Chandra Source Catalog \citep{evans2010}, and so we exclude this source as well, leaving 57 dwarf galaxies which do not exhibit evidence for AGNs. While we do not find evidence for AGN activity in these 57 dwarf galaxies, we cannot definitively rule out the presence of low-luminosity AGNs that remain undetected. 

To demonstrate that these non-AGN-hosting dwarf galaxies have similar physical properties as the comparison AGN hosts, Figure \ref{fig:grcolormass} shows the $g - r$ color-mass diagram of this sample, as well as the \citet{kimbrell2021} sample for comparison. Figure \ref{fig:sizemass} shows the size-mass diagram, using the Petrosian 50\% light radius reported by the NSA, of the two samples for comparison.

\section{Analysis}\label{sec:analysis}

The aim of this work is to analyze the structures of dwarf galaxies that do not show evidence of hosting an active massive BH, so that we may compare these results to those of galaxies that do show evidence of hosting an active massive BH. We are especially interested in finding how often these galaxies possess an inner (pseudo)bulge component in addition to a disk, and whether the (pseudo)bulge or disk dominates the total galaxy light. Irregular galaxies make up 14\% (8/57) of our sample and their structures can not be accurately modeled in GALFIT \citep{peng}. 
For the 49 dwarf galaxies in our sample with regular morphologies, we model each galaxy with four models using one or two S{\`e}rsic components with and without point spread functions (PSFs). The PSFs may represent nuclear star clusters (NSCs), which are known to be common in galaxies in this mass range \citep{carlsten2021,sanchezjanssen2019,neumayer}. The inner S{\`e}rsic component of a two component model can be interpreted as the (pseudo)bulge, with the outer component fixed as an exponential disk. For galaxies that are viewed edge-on in our imaging, we replace the S{\`e}rsic model representing the disk with an edge-on disk model.

\subsection{PSF Construction}

\begin{figure}[!h]
\begin{center}
\includegraphics[width=3.25in]{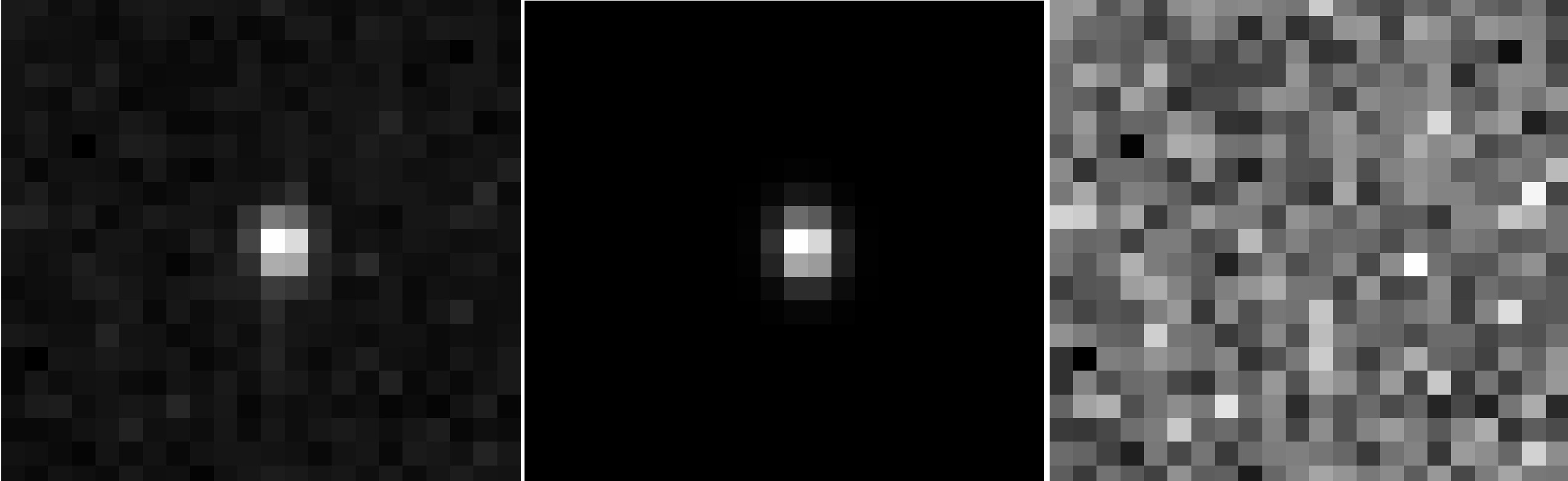}
\end{center}
\caption{Star used for PSF creation for Galaxy UK6 (left); model PSF created using GALFIT (middle); residuals resembling random noise after subtracting the model PSF from the image of the star (right).}
\label{fig:psfexamples}
\end{figure}

An important step in modeling these galaxies is construction of an accurate PSF, which models the detector response to a point source of light. An inaccurate PSF risks inaccurate modeling of the galaxy as a whole. The variability of seeing conditions requires a PSF to be created for each galaxy. For each galaxy, we selected an isolated star and cut out a $\sim$ 20 $\times$ 20 pixel square centered on the star. We then created a model of the star in GALFIT using as many S{\'e}rsic components as necessary until the residuals showed only random noise. This is a standard method for PSF creation using GALFIT\footnote{users.obs.carnegiescience.edu/peng/work/galfit/TFAQ.html}. We do not attempt to model physical parameters of the starlight, so the number of S{\'e}rsic components and their values are not important. We simply require an analytic model that accurately captures the shape of the star. We subtract the estimated sky background in order to create a model of the star with a very high signal-to-noise ratio, as required of PSFs. Figure \ref{fig:psfexamples} shows an example of a star image, the corresponding PSF model and the residuals.

We performed fitting on each galaxy with inaccurate PSFs as well, testing the PSF fitting as a whole. The use of an inaccurate PSF will lead to systematic structures in the residuals of every attempted model. This will most often occur if the background of the PSF is not perfectly subtracted, or if an improper model (such as a pure Gaussian) is used. In the former case, a bright square will appear in the center of the galaxy upon modeling. In the latter case, the residuals will show obvious signs of subtracting a symmetric model rather than a typically somewhat asymmetric star profile.

\subsection{Galaxy Modeling}
We used the galaxy fitting software GALFIT \citep{peng} to fit analytical two-dimensional models to our images. There are many models that GALFIT can use to fit galaxies, many of which are used for specific morphologies (e.g. the de Vaucouleurs profile which is often used to model classical bulges \citep{devauc}, or the edge-on disk function). For the galaxies in this work, we attempted to use the very general S{\'e}rsic profile, which takes the form \citep{sersic}:

\begin{equation}
  \Sigma(r) = \Sigma_e \times \textrm{exp}\bigg[-\kappa\Big(\Big(\frac{r}{r_e}\Big)^{\frac{1}{n}}-1\Big)\bigg]
\end{equation}

\noindent
where $r_e$ is the effective radius such that half of the flux lies within $r_e$; $\Sigma_e$ is the surface brightness at the effective radius $r_e$. The parameter $n$ is the S{\'e}rsic index and is coupled to the parameter $\kappa$. A higher S{\'e}rsic index indicates extended wings and a sharp brightness profile at the center of the galaxy. $\kappa$ is defined as $\kappa = {\Gamma(2n)}/{2}$, with $\Gamma$ being the complete gamma function. The case of $n = 4$ reduces to the de Vaucouleurs profile, $n = 1$ corresponds to an exponential disk profile, and $n = 0.5$ is a Gaussian profile.

We followed the same fitting process as \citet{kimbrell2021}, recommended by \citet{peng}. We began by fitting a single S{\`e}rsic component to a given galaxy, which in some cases was the best fitting model. However, even if it was a poor fit, this model gave us basic structural information about the galaxy (e.g. effective radius, axis ratio) that informed the initial conditions of more complex models. Following the single-S{\`e}rsic model, we added a PSF and re-ran GALFIT, using the results of the previous run as the initial parameters for the S{\`e}rsic component. This PSF could represent a nuclear star cluster, which are known to be common in galaxies in this mass range \citep{neumayer}. 

We proceeded with our modeling by fitting the galaxies with a two-S{\`e}rsic model with no PSF. We let every parameter of the inner S{\`e}rsic component vary while fixing the S{\`e}rsic index $n$ of the outer S{\`e}rsic component at $n=1$, which is the typical value for an exponential disk. Even in galaxies which clearly have a bright point source at their center, starting with a simpler model without a PSF is important in order to get structural information about the individual components before adding more complexity with a PSF. Finally, we fit a model with an inner S{\`e}rsic component, an exponential disk and a PSF.

Every model previously discussed also includes a sky component. We assume a flat sky with no variations across the whole of each image, and we let GALFIT fit for the brightness of the sky. In principle, a small patch of the sky that is significantly brighter or dimmer than the rest can be masked out, allowing GALFIT to ignore it in the fitting. However, this was never necessary for this work, and a flat sky adequately modeled the background light in every instance.

Unlike the \citet{kimbrell2021} sample, this sample includes several galaxies with regular morphologies that are not well fit by any of the above models. Some galaxies are edge-on and there are also some spiral galaxies. When possible, these were modeled using the more specific GALFIT components, following the same basic path detailed above: first the simplest model (edge-on disk, spiralled S{\`e}rsic, etc.) was applied without a PSF, then a PSF was included, then any other components were added as necessary (e.g., a bulge inside the disk, a non-spiralled interior component). However, while spirals can be very obvious to the eye, GALFIT can struggle to accurately distinguish a spiral. This will not significantly affect the results of the decomposition that we perform, and so galaxies with spirals that cannot be modeled well by GALFIT were modeled in the same way as galaxies without spirals, with acceptance that some spiral structure will be present in the residuals.

\begin{figure*}[!h]
\includegraphics[width=\textwidth]{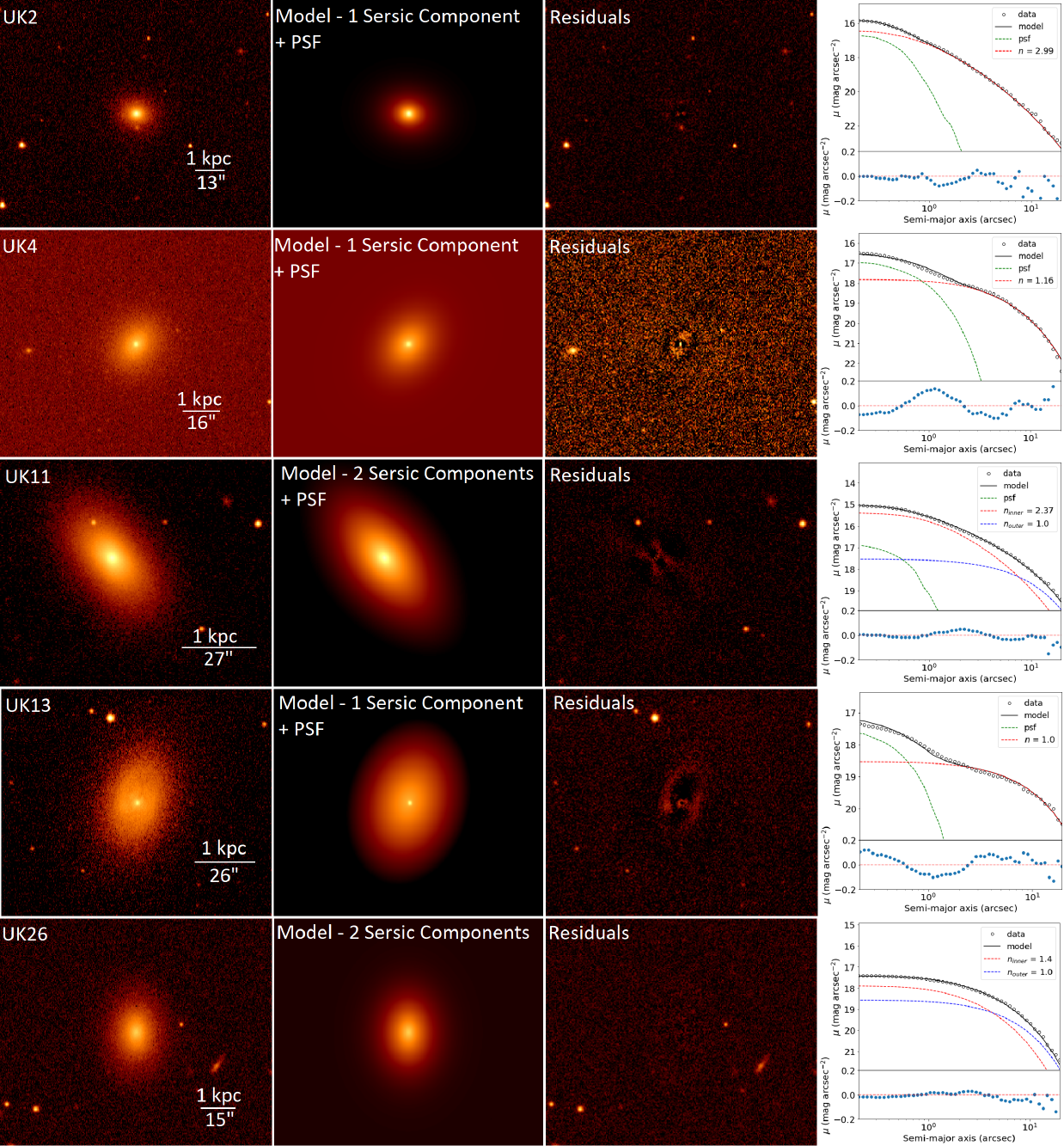}
\caption{Left: {UKIRT image, GALFIT model, and the residuals obtained by subtracting the GALFIT model from the image. Images are shown on a log scale, and the residuals are very stretched to show faint details. Right: Surface brightness profiles. Data are black circles and the model is shown as a black line; the GALFIT components are shown as dashed lines in color. The bottom panels contain the radial residuals.}}
\label{fig:results}
\end{figure*}

\subsection{Model Selection}\label{sec:model}
With the analysis completed, we determined which model best fit each galaxy. We simultaneously considered all possible models for each galaxy in order to pick the best one. In 26 of the 49 galaxies we modeled, GALFIT was unable to converge on any two component model, leaving only models including one S{\`e}rsic with or without a PSF. In 13 others, GALFIT would only converge on a two component model with non-physical parameters, such as a (pseudo)bulge that is dominant over the disk everywhere, an unrealistically high S{\`e}rsic index, or an effective radius indicative of the S{\`e}rsic component attempting to fit the sky background. For the 8 galaxies in which GALFIT converged to a high inner S{\`e}rsic index, we re-attempted the modeling with the inner S{\`e}rsic index constrained to $n \leq 5$, but these runs never resulted in a better fit than the non-constrained runs. 

For model selection, we followed the process of \citet{oh2017} to be consistent with \citet{kimbrell2021}. This was a three-step selection process. First, we examined the radii of both components in the two component models. If the exponential disk had an effective radius smaller than that of the inner (pseudo)bulge, that two component model was rejected. Next, we examined the inner component. If a two component model led to the inner S{\`e}rsic being dominant over the disk at all radii, we rejected that two component model. The inner component was also our primary metric when deciding whether or not to include a PSF. A galaxy that requires a PSF, yet one is not included in the model, will often have a high inner S{\`e}rsic component to try to capture the excess of light at the center. Evidence from observations of dwarf galaxies with and without AGNs (see, e.g. \citealt{schutte, jiang, coma2003, kormendy}) supports both dwarf ellipticals and (pseudo)bulges having S{\`e}rsic indices $n \leq 4$.  In order to have a conservative rejection criterion, we rejected any PSF-included model that included an inner S{\`e}rsic component $n \geq$ 5. Examination of the residuals also helped select for PSF necessity. A S{\`e}rsic component attempting to account for an inner point source leads to telltale concentric circles in the residuals, alternating light and dark, near the galaxy center. If visual inspection of the residuals show these rings when the PSF is absent, we took this as evidence of the necessity of a PSF component.
The final selection metric between the models was the Akaike Information Criterion (AIC). The AIC is a statistical criterion used to select between different models, and can be calculated from $\chi ^2$, assuming normally distributed noise, as follows:

\begin{equation}
{\rm AIC} = \chi ^2 + 2k
\end{equation}

\noindent with $k$ being the number of free parameters in a given model. A more complicated model (i.e., one with an additional component) must improve the AIC by at least 10 to be accepted over the less complicated one, to remain consistent with \citet{kimbrell2021}. Of the 49 dwarf galaxies with regular morphologies, 36 were best fit by a single component model, leaving 13 best fit by a two component model. A PSF was required for 41 of the 49 regular dwarf galaxies. These results are further discussed in Section \ref{sec:results}.

\section{Results and Discussion}\label{sec:results}

\subsection{Demographics of Dwarf Galaxy Sample}

\begin{figure}[!h]
\centering
\includegraphics[width=3.25in]{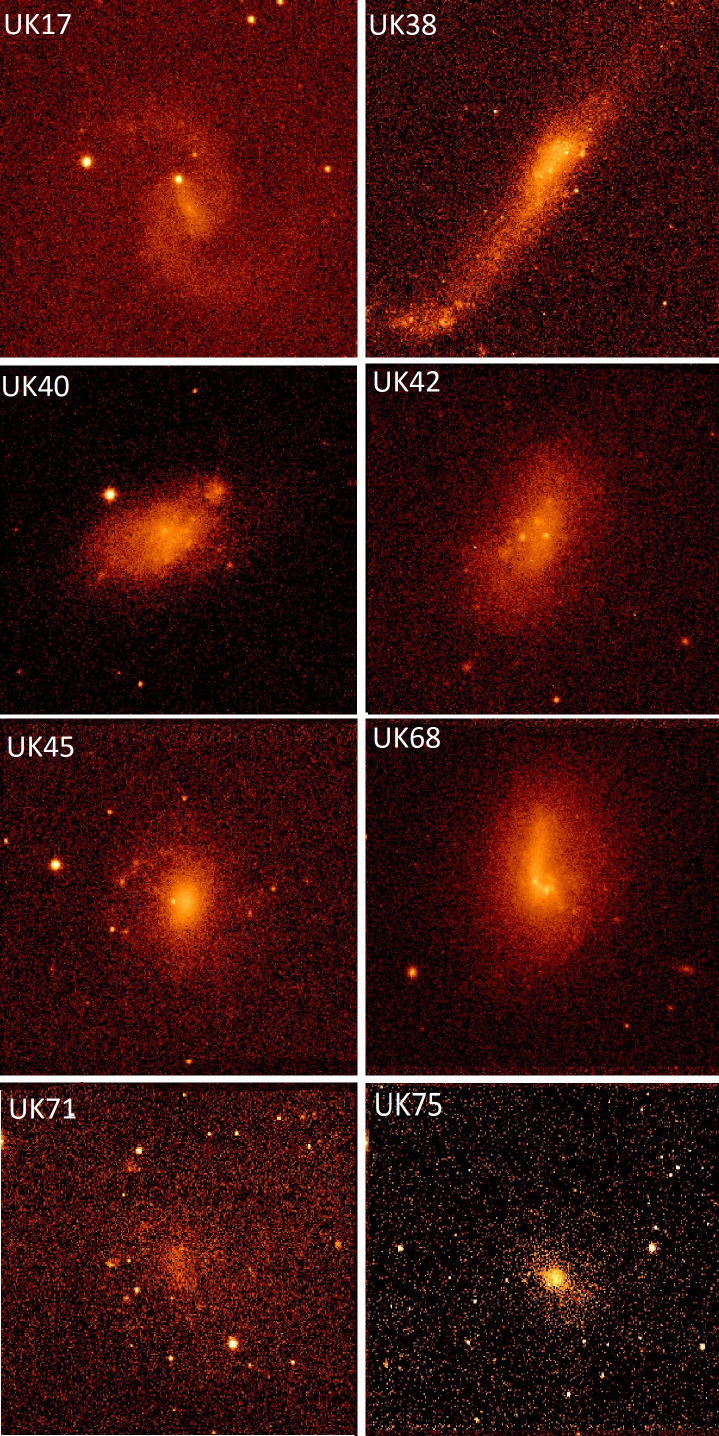}
\caption{UKIRT images of the 8 irregular/disturbed galaxies in our sample. All images are shown on a log scale.}
\label{fig:irregulars}
\end{figure}

Of our sample of 57 non-AGN-hosting dwarf galaxies studied here, 49 galaxies have regular morphologies (86\%). We modeled these in GALFIT using either one or two S{\`e}rsic components, sometimes with a PSF, or an edge-on disk (and possibly a bulge). Figure \ref{fig:results} shows some examples and we report the results of our modeling in Table \ref{table:faceonresults}. Meanwhile, 8/57 (14\%) have irregular/disturbed morphologies and could not be accurately modeled in GALFIT (see Figure \ref{fig:irregulars}). 

The majority (39) of the regular dwarf galaxies are relatively face-on, while 10 are edge-on dwarfs. Of the 39 face-on dwarfs, we have determined that 30 ($\sim 78\%$) are best fit by a single S{\`e}rsic component, while nine ($\sim 22\%$) are best fit by a two-component (pseudo)bulge + disk model.  For the galaxies best fit by a single S{\`e}rsic component, we differentiate between elliptical/(pseudo)bulge-like galaxies and disk-like galaxies using the S{\`e}rsic index $n$. A dwarf galaxy fit by a component with $n < 1.5$ is called ``disk-like", to retain consistency with \citet{kimbrell2021}. A dwarf fit by a component with $n \geq 1.5$ is called ``elliptical" or ``(pseudo)bulge-like". Of the 10 edge-on dwarf galaxies, 4 are best fit with a disk + a (pseudo)bulge, while the other six are best fit with a single-component edge-on disk. We find that a PSF component is required in $\sim$ 84\% of these regular galaxies. 

In order to gauge how well this sample represents the general population of dwarf galaxies, we compare our results to dwarf galaxy demographics explored by \citet{reines2022} using the Catalog and Atlas of the Local Volume Galaxies \citep{karachentsev}. Assuming $K$-band luminosity is a good proxy for stellar mass to first order, Table 1 in \citet{reines2022} indicates that 63\% of galaxies in the mass range $10^{9} < {M_*}/M_\odot < 10^{9.5}$ are late-type spiral galaxies (Sdm, Sd) while 26\% are irregular galaxies, Magellanic irregulars and blue compact dwarf galaxies. \citet{reines2022} quotes a total of 12\% for earlier-type galaxies in this mass range.

\begin{figure}[!h]
\hspace{-.2cm}
\includegraphics[width=\linewidth]{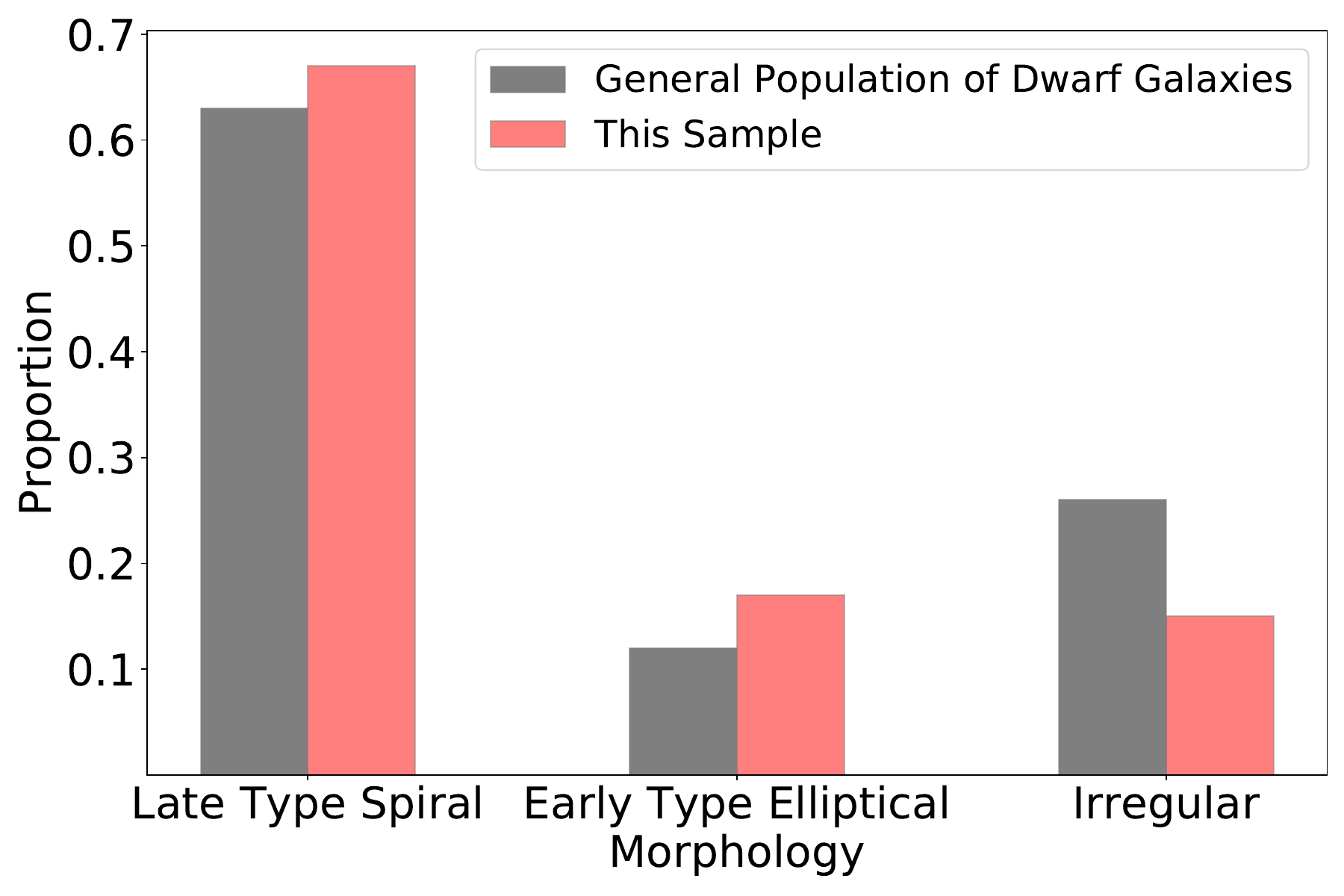}
\caption{Morphologies of the galaxies analyzed in this work (red) compared to the general population of dwarfs in the same mass range ($10^{9} < {M_*}/M_\odot < 10^{9.5}$) discussed in \citet{reines2022} (black). This demonstrates that our sample of non-AGN hosting dwarf galaxies provides a good representation of the general population of dwarf galaxies.}
\label{fig:reinescomparison}
\end{figure}

While not identical, we find comparable statistics among our sample of dwarf galaxies without evidence of active massive BHs studied here, which spans the mass range given above  -- $\sim$68\% of the galaxies studied here are either pure disks ($\sim46$\%) or disk-dominated galaxies with fainter (pseudo)bulges ($\sim$22\%). Irregular galaxies make up 14\% of our sample and $\sim 18\%$ of the galaxies are best fit with elliptical/spheroidal models. Overall, we find a higher proportion of early-type galaxies than reported by \citet{reines2022} and fewer irregulars, but as a whole our sample is fairly consistent with the general population of nearby dwarf galaxies. We show the comparison in Figure \ref{fig:reinescomparison}.

We also compare our sample to the that analyzed in \citet{gama2012}. In that work, a sample of 2711 galaxies in the local universe (0.025 $<$ z $<$ 0.06) were selected from the Galaxy and Mass Assembly Survey (GAMA). These galaxies were separated by their morphologies and binned by total stellar mass. It was found that at low stellar mass, ${\rm log (M_*/M_\odot)} \lesssim 10$, the local universe is mostly populated by disk-dominated and irregular galaxies, with a low fraction of elliptical and spheroid-dominated galaxies, further confirming that our sample of inactive galaxies is representative of the local volume of galaxies in this mass range.

\subsection{Comparison to Dwarf Galaxies with AGNs}

First, we note that the fraction of regular (86\%) and irregular (14\%) galaxies found here for the non-AGN-hosting dwarf galaxies is very nearly the same as for the AGN-hosting galaxies presented in \citet{kimbrell2021} (85\% and 15\%, respectively). Below, we focus on the similarities/differences between dwarf galaxies with regular morphologies in the two samples.

\begin{figure}[!h]
\hspace{-.2cm}
\includegraphics[width=\linewidth]{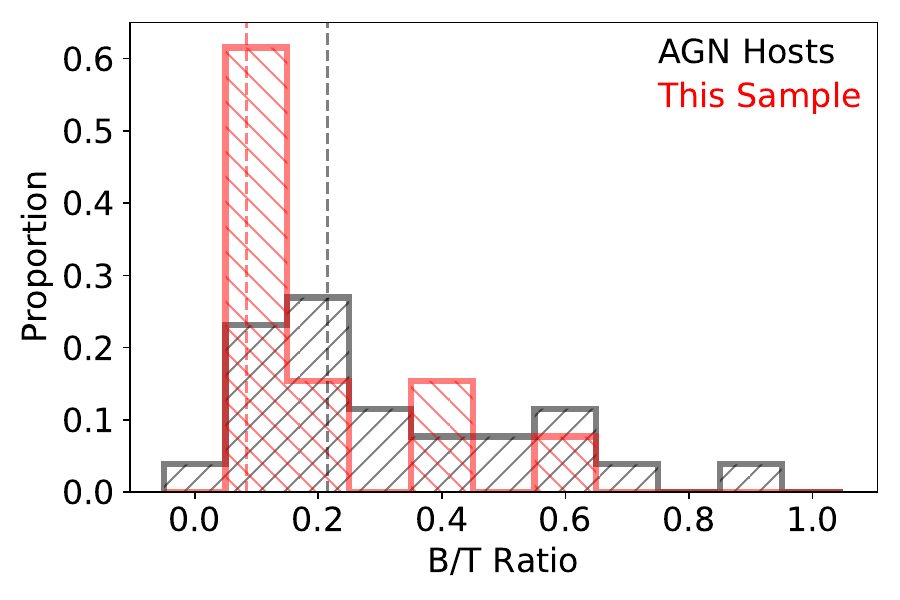}
\caption{B/T ratios of the 2-component galaxies in this work (red) compared to the 2-component AGN hosts analyzed in \citet{kimbrell2021} (black). Dashed vertical lines indicate median values.}
\label{fig:btratios}
\end{figure}

The dwarf galaxies in our sample best fit by two S{\`e}rsic components, indicating a (pseudo)bulge + disk, have bulge to total light ratios somewhat lower on average than the AGN hosts studied in \citet{kimbrell2021}. Here, the two-component galaxies have a median bulge-to-total ratio (with PSF subtracted) $<$B/T$>$ = 0.08, compared to 0.21 for the AGN hosts. We also find a much smaller proportion of our current sample requiring a two component model; only $\sim 23\%$ of our face-on galaxies are so modeled ($\sim 27\%$ when including edge-on galaxies), while $\sim 76\%$ of galaxies modeled in \citet{kimbrell2021} had a (pseudo)bulge and a disk. The presence of edge-on galaxies in our sample also differentiates it from the \citet{kimbrell2021} sample. We report 10 ($\sim 18\%$) of the galaxies in our sample are edge-on. \citet{kimbrell2021} reported no edge-on galaxies at all. However, this may be a selection effect since low-luminosity AGNs may be more difficult to detect in edge-on galaxies where the light can be more easily obscured.

We apply the two-sample Kolmogorov-Smirnov (K-S) test \citep{massey1951} to the distribution of B/T ratios of this sample of galaxies and the B/T ratios of the AGN hosts modeled in \citet{kimbrell2021}. The two-sample K-S test takes in two samples of data and returns a statistic which measures the maximum distance between the two empirical cumulative distribution functions. The benefit to the two-sample K-S test is that it can help investigate whether two samples are likely to be drawn from the same underlying distribution without needing to know that distribution. For these two samples of B/T ratios, the two-sample K-S test returns a $\sim$ 0.8\% chance of drawing data as dissimilar as these if they came from the same distribution of B/T ratios. However, we also run the two-sample K-S test comparing just the two-component galaxies in the AGN host sample and our sample. This way, we compare the (pseudo)bulge + disk galaxies that were selected as AGN hosts to the (pseudo)bulge + disk galaxies that were not. In this case, we find that there is a $\sim 4\%$ chance of drawing B/T ratio distributions so dissimilar if they came from the same distribution. It seems that while some difference exists structurally between the AGN hosts in \citet{kimbrell2021} and the sample presented here, the primary difference between them is the presence (or lack thereof) of a (pseudo)bulge.

\begin{figure}[!h]
\hspace{-.2cm}
\includegraphics[width=\linewidth]{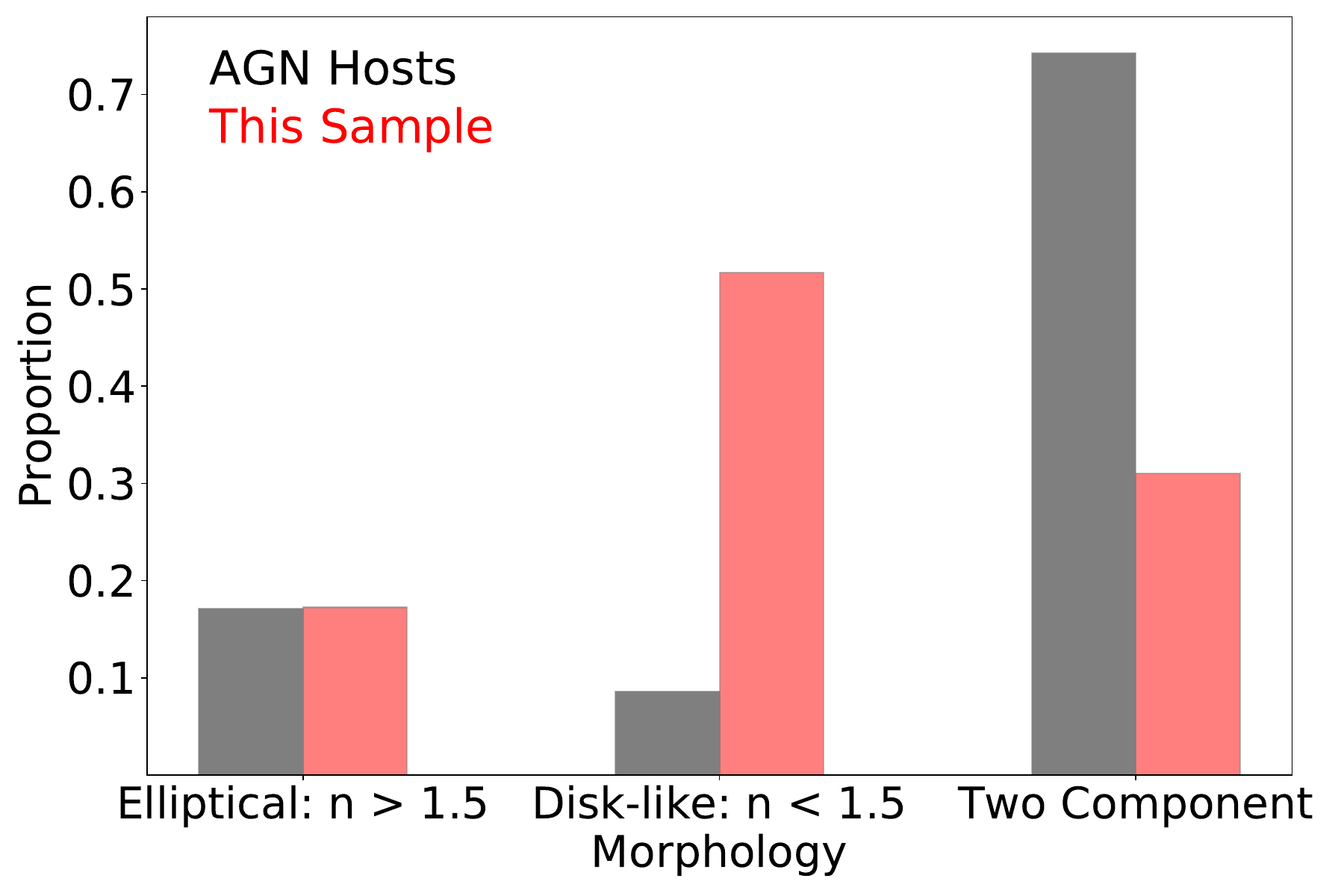}
\caption{Morphologies of the galaxies analyzed in this work (red) compared to the AGN hosts analyzed in \citet{kimbrell2021} (black). The first and second options indicate a single component galaxy best fit as a dwarf elliptical or a pure disk, respectively. The third option indicates a dwarf galaxy best fit with two S{\`e}rsic components.}
\label{fig:morphcomparison}
\end{figure}

We further show this in Figures \ref{fig:btratios} and \ref{fig:morphcomparison}. In Figure \ref{fig:btratios} we show the distribution of B/T ratios for strictly the two-component galaxies in each sample. We find a somewhat similar distribution, with the majority of each sample of two-component dwarf galaxies being disk-dominated. In Figure \ref{fig:morphcomparison} we show the number of each morphology type present in each sample. While both the AGN hosts of \citet{kimbrell2021} and the dwarf galaxies in this sample show similar fractions of dwarf ellipticals, the sample analyzed in this work is dominated by disk-like single-component galaxies, which are hardly present in \citet{kimbrell2021}. This reinforces the notion that the primary structural difference between this sample and the AGN hosts analyzed in the previous work is the presence of a (pseudo)bulge, rather than the relationship between the disk and (pseudo)bulge, when present.

\begin{figure}[!h]
\hspace{-.2cm}
\includegraphics[width=\linewidth]{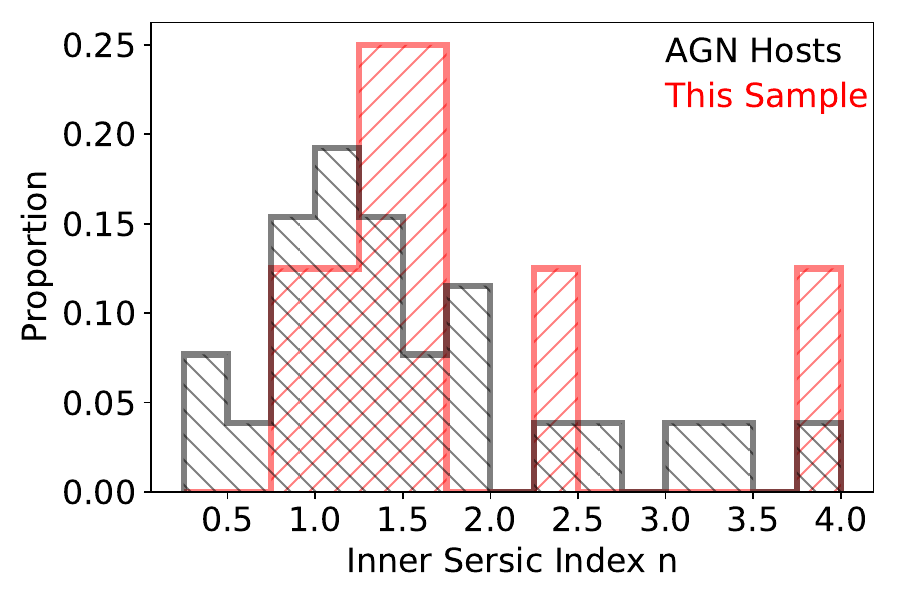}
\caption{Inner S{\`e}rsic indices of all 2-component galaxies studied in this work (red) compared to the AGN hosts analyzed in \citet{kimbrell2021} (black)}.
\label{fig:innerindices}
\end{figure}

We also show the S{\`e}rsic indices of the inner components of our two-component galaxies in Figure \ref{fig:innerindices}. We find that the inner (pseudo)bulge components of our sample were structurally similar to the \citet{kimbrell2021} sample of AGN hosts; our galaxies had a median inner S{\`e}rsic index $<n> = 1.49$, compared to $<n_{\rm AGN Hosts}> = 1.31$. Performing the two-sample K-S test on the inner S{\`e}rsic indices supports this similarity, finding a $\sim$ 61\% chance of drawing data this dissimilar if they came from the same distribution of S{\`e}rsic indices. We reiterate that only nine of the galaxies in our sample possessed a (pseudo)bulge, so this comparison must be taken with extra caution.

\subsection{Nature of the Point Sources}\label{sec:pointsource}
Despite not being selected as hosts of AGNs, 41 of the 49 ($\sim$ 84\%)
regular galaxies in our sample are best fit by a model including a point source of light. As shown in the literature  \citep[e.g.,][]{neumayer,sanchezjanssen2019,denBrokComa,kimbrell2021}, this is entirely consistent with the proportion of galaxies in this mass range which are expected to host nuclear star clusters (NSCs).

\begin{figure}[!h]
\hspace{-.2cm}
\includegraphics[width=\linewidth]{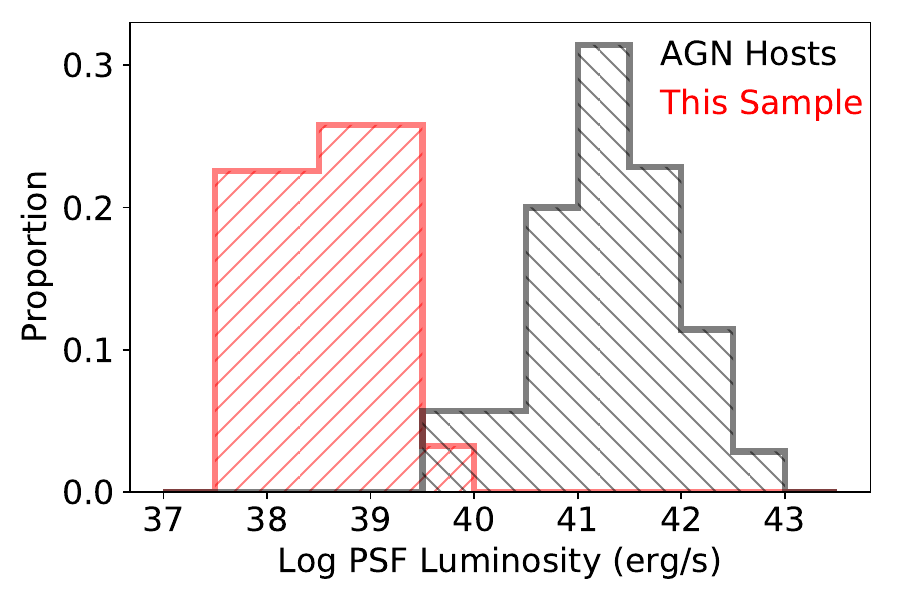}
\caption{Log luminosity in ergs/s of the central PSF components of this sample (red) and the \citet{kimbrell2021} sample of AGN hosts (black). The PSFs in the AGN hosts are, on average, more than two orders of magnitude more luminous than those in the non-AGN hosts, suggesting the PSFs in the AGN hosts are dominated by AGN light.}
\label{fig:psflumcompare}
\end{figure}

We search for any difference between the PSFs of this inactive sample and the PSFs of the AGN-hosting dwarf galaxies of \citet{kimbrell2021}. We show the comparison between the luminosities in Figure \ref{fig:psflumcompare}, which shows a large gap between the two samples. On average, the PSFs in the AGN-hosting dwarf galaxies are 310 times more luminous than those in the non-AGN-hosting dwarfs. While we cannot definitively determine the origin of the point sources in these galaxies, the marked difference between the PSF luminosities in the two samples suggests that the PSFs in the \citet{kimbrell2021} sample are dominated by AGN light, while the proportion of galaxies in this inactive sample hosting a PSF in the best-fit model is consistent with the proportion of galaxies in this mass range hosting nuclear star clusters \citep{neumayer}.

\begin{figure}[!h]
\hspace{-.2cm}
\includegraphics[width=\linewidth]{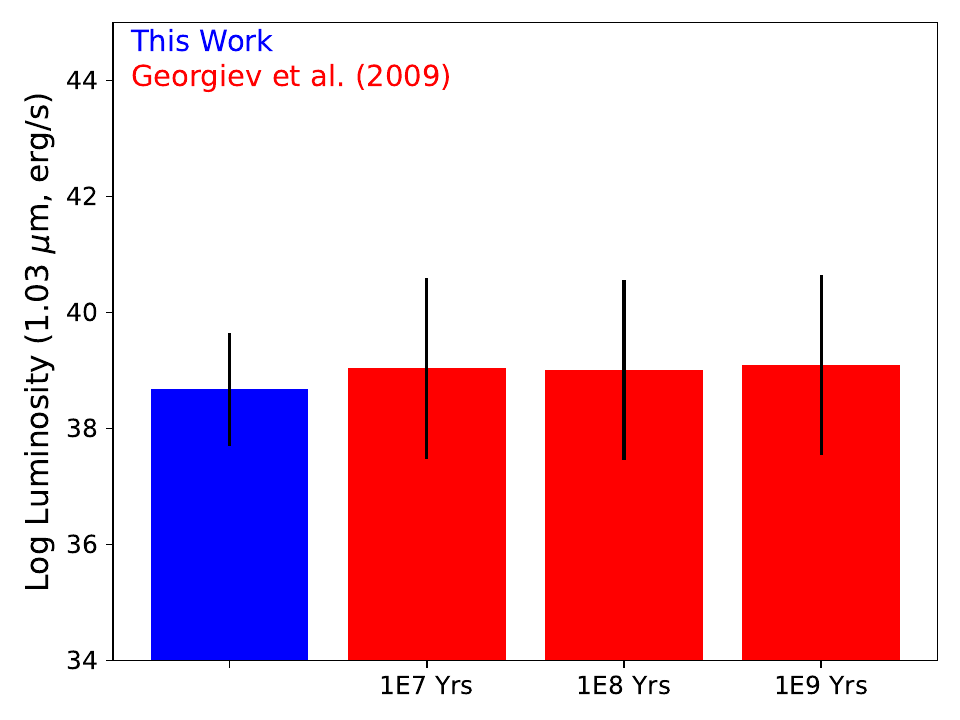}
\caption{Median log luminosity in erg/s for PSFs of this sample (blue bar) and the median predicted $1.03 \mu m$ luminosity for the PSFs of the \citet{georgiev2009} sample (red bars) at varying ages. Black lines indicate the range from minimum to maximum luminosity.}
\label{fig:nscgeorgiev}
\end{figure}

We compare the $1.03 \mu$m luminosities of the PSFs in our sample to the nine nuclear star clusters in late-type dwarf galaxies studied in \citet{georgiev2009}. We use the Python synphot routine to convert the luminosities reported there in the ACS F814W filter to $1.03 \mu$m using \citet{bruzualcharlot2003} population synthesis models at a variety of ages, assuming no extinction.  We find that the luminosities of our PSFs are consistent with the estimated luminosities of known nuclear star clusters in dwarf galaxies (see Figure \ref{fig:nscgeorgiev}).

\begin{figure}[!h]
\hspace{-.2cm}
\includegraphics[width=\linewidth]{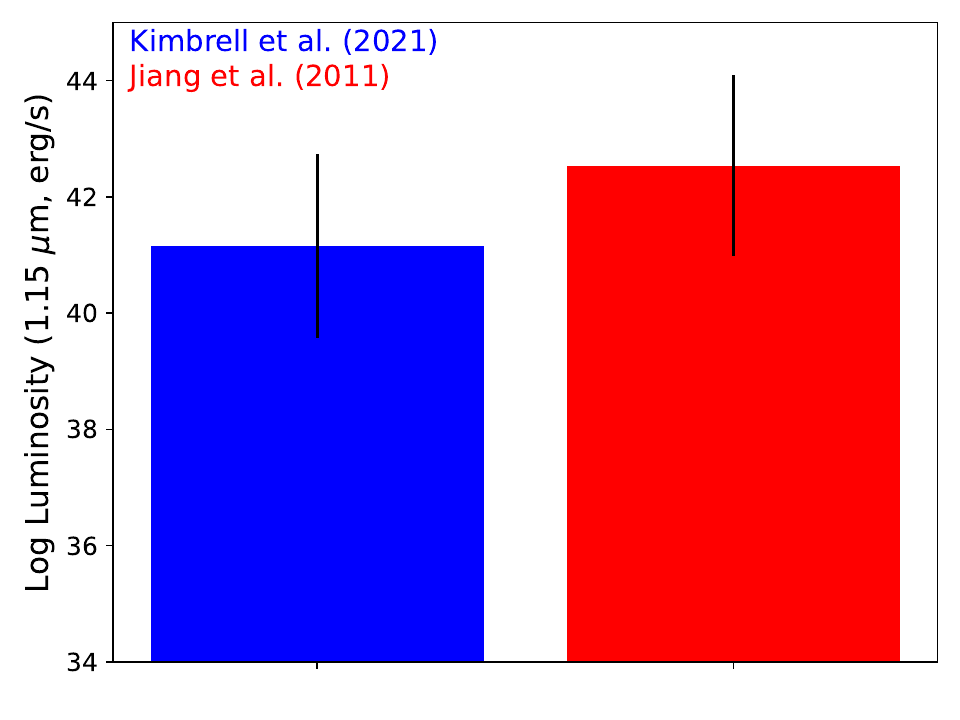}
\caption{Median log luminosity in erg/s for the PSFs of the \citet{kimbrell2021} (blue bar) and the median predicted $1.15 \mu m$ luminosity for the PSFs of the \citet{jiang} sample (red bar). Black lines indicate the range from minimum to maximum luminosity.}
\label{fig:agnjiang}
\end{figure}

We perform a similar comparison between the PSFs of the \citet{kimbrell2021} AGN hosts and the PSFs of the hosts of low-mass black holes reported by \citet{greeneho2007} and studied by \citet{jiang} in the WFPC2 F814W filter. We use a power law spectrum for an AGN: $f_\lambda \propto \lambda^\alpha$ with $\alpha = -1.56$ when $\lambda \leq 5000$ \AA and $\alpha = -0.45$ for $\lambda > 5000$ \AA \citep{vandenberk2001}. Using this spectrum, we convert the luminosities reported in \citet{jiang} for the PSFs to the F110W (central wavelength 1.15 $\mu$m) filter used in \citet{kimbrell2021}. The \citet{greeneho2007} sample are more massive galaxies than the dwarfs studied in \citet{kimbrell2021} and host more massive BHs than the dwarfs. In addition, most of the dwarfs do not exhibit broad emission lines like the \citet{greeneho2007} sample, which may indicate more obscured AGNs in \citet{kimbrell2021}. For these reasons, we do not expect the \citet{kimbrell2021} PSFs to be as luminous. We find that the dwarf AGNs hosts have PSF luminosities lower than the \citet{greeneho2007} sample; however, given the considerations above, these are more consistent with the more massive AGNs than the nuclear star clusters found in dwarf galaxies. This suggests that the PSFs found in the sample of non-AGNs studied here are likely NSCs, while the PSFs in \citet{kimbrell2021} are likely dominated by AGN light. We show the results of both of these comparisons in Figures \ref{fig:nscgeorgiev} and \ref{fig:agnjiang}, on identical vertical axes for ease of comparison.

While the sample of dwarf galaxies studied in this work have PSFs consistent in luminosity with NSCs, we cannot definitively rule out contributions from low-luminosity AGNs with these data alone. Indeed, \citet{neumayer} discuss the evidence for the presence of massive BHs within NSCs. However, we find no evidence for AGNs in this sample as described in Section \ref{sec:sample}. Moreover, as discussed above, several pieces of evidence indicate the PSFs are dominated by light from NSCs rather than low-luminosity AGNs. First, we find the luminosities are consistent with previously observed NSCs in dwarf galaxies. Second, the PSFs appear in a fraction of this sample which is consistent with the occurrence of NSCs in galaxies in this mass range. Finally, the range of luminosities of the PSFs in this sample are nearly completely distinct from the luminosities of the PSFs in the AGN sample of \citet{kimbrell2021} (Figure \ref{fig:psflumcompare}).

\section{Conclusions}\label{sec:conclusion}

We presented a study of the morphologies and structures of 57 dwarf galaxies that do not show signs of hosting AGNs. We have compared the demographics of this sample to the study performed on dwarf galaxies that host BPT-selected AGNs in \citet{kimbrell2021}. We summarize our results below:

\begin{enumerate}

  \item The sample of non-AGN-hosting dwarf galaxies analyzed in this work, selected to span the same range of stellar mass and physical resolution as the \citet{kimbrell2021} sample of AGN-hosting dwarfs, is representative of the general population of dwarf galaxies in the local volume (see Figure \ref{fig:reinescomparison}).

  \item Overall, we find nearly the same fraction of regular (86\%) versus irregular (14\%) galaxies among the dwarf galaxy samples with and without optically-selected AGNs. 
  
  \item  Both samples also have similar fractions of regular, early-type/elliptical dwarf galaxies ($\sim18\%$).

  \item The primary morphological difference between the two samples is among the galaxies with disks. A large fraction of the non-AGN hosts are pure disk galaxies with no detectable (pseudo)bulges, while the vast majority of the AGN-hosting dwarfs have detectable (pseudo)bulges (see Figure \ref{fig:morphcomparison}).

  \item We also find 10 ($\sim 18\%$) edge-on galaxies in our sample here, of which none were found among the AGN hosts. This could be a selection effect since low-luminosity AGNs are likely more difficult to detect in edge-on galaxies where the light can be more easily obscured.

  \item A central point source of light is present in $\sim 84\%$ of the best-fit models for the regular dwarf galaxies in the sample of non-AGN hosts (excluding edge-on galaxies). This proportion is consistent with the expected fraction of galaxies hosting nuclear star clusters in this stellar mass range \citep{neumayer}. We find the luminosities of these point sources to be consistent with estimated luminosities at this wavelength of known NSCs in dwarf galaxies \citep{georgiev2009}.

  \item In contrast, all of the AGN host galaxies  require a point source in their best fit model and the average luminosity of the PSFs is more than two orders of magnitude higher than that of the non-AGN hosts, and are largely consistent with luminosities expected of AGNs in this wavelength \citep{jiang,greeneho2007}, suggesting the point sources in the AGN hosts are dominated by AGN light.  
  
\end{enumerate}

We have shown that dwarf galaxies with BPT-selected AGNs are overall structurally different from the general population of dwarf galaxies. Primarily, we have found that dwarf galaxies hosting AGNs are more likely to host a central pseudobulge than the general population of dwarf galaxies, which contains a much higher fraction of pure disks. While the BPT diagram certainly does not capture all of the AGNs in dwarf galaxies \citep{reines2022}, our results agree with the study of \citet{satyapal2009} who found that AGNs in pure bulgeless galaxies are indeed rare in this mass range based on searching for high-ionization emission lines using mid-IR spectroscopy. It is also notable that the pure disk galaxy M33 does not appear to have a nuclear BH based on a dynamical upper limit of $\sim 1500~M_\odot$ \citep{gebhardt2001}.

It is known that BHs do not correlate with pseudobulges in the same way that BHs tightly correlate with classical bulges and elliptical galaxies \citep[e.g.,][]{kormendy2011}. In the latter, major mergers are thought to be responsible for driving large amounts of gas to galaxy centers that can quickly grow both the bulge and the BH. In contrast, pseudobulges grow slowly via inward gas transport and their BHs evolve independently, growing as low-level Seyferts via local and stochastic processes. Indeed, the AGN-hosting dwarfs in \citet{kimbrell2021} are dominated by disk galaxies with pseudobulges and the BHs are detected as Seyfert nuclei. However, we do not know if BHs are preferentially {\it formed} in dwarfs with pseudobulges or if gas is more effectively funneled to the nucleus in dwarfs with pseudobulges making their BHs detectable as AGNs.

In any case, our findings indicate that AGNs are likely to be preferentially found in dwarf galaxies containing (pseudo)bulges rather than pure disks. 
This has implications for the search for BHs in dwarf galaxies, which is important for constraining the BH occupation fraction in low mass galaxies. The BH occupation fraction in dwarfs, in turn, is a key diagnostic for discriminating between  possible seeding mechanisms of supermassive black holes \citep{volonteri, Greene, ricarte2018}.

\acknowledgements
A.E.R. gratefully acknowledges support for this paper provided by NASA through EPSCoR grant No. 80NSSC20M0231. This work is based in part on data obtained as part of the UKIRT Infrared Deep Sky Survey. We thank Chien Peng and the GALFIT help forum for helpful discussions on using GALFIT.
\appendix

\LongTables
\begin{deluxetable*}{ccccccc}[!h]
\tabletypesize{\footnotesize}
\tablewidth{0pt}
\tablecaption{Fitting Results for Face-on Galaxies With Regular Morphologies \label{table:faceonresults}}
\tablehead{
\colhead{ID} & \colhead{Component} & \colhead{\textit{m}$_{Y}$} & \colhead{n} & \colhead{$R_{e}$ (kpc)} & \colhead{q} & \colhead{Additional Components} \\
\colhead{(1)} & \colhead{(2)} & \colhead{(3)} & \colhead{(4)} & \colhead{(5)} & \colhead{(6)} & \colhead{(7)}}
\startdata
& PSF & 22.95 & - & - & - \\
UK1 & S{\'e}rsic & 16.18 & 1.59 & 3.50 & 0.38 \\

\\
& PSF & 20.02 & - & - & - \\
UK2 & S{\'e}rsic & 16.50 & 2.99 & 0.43 & 0.83 & \\

\\
& PSF & 18.99 & - & - & - \\
 UK4 & S{\'e}rsic & 15.64 & 1.16 & 0.12 & 0.78\\
 
  \\

& PSF & 21.04 & - & - & - \\
UK6 & S{\'e}rsic & 16.27 & 1.45 & 0.61 & 0.94 \\
\\

& PSF & 20.44 & - & - & - \\
UK10 & S{\'e}rsic & 15.84 & 1.01 & 0.97 & 0.89 \\
\\
\\
\\

& PSF & 19.70 & - & - & - \\
UK11 & Inner S{\'e}rsic & 14.43 & 2.37 & 0.63 & 0.73 \\
& Outer S{\'e}rsic & 14.79 & 1.00 & 1.11 & 0.47\\
\\

& PSF & 21.43 & - & - & - \\
UK12 & S{\'e}rsic & 15.20 & 1.76 & 0.88 & 0.44 \\
\\

& PSF & 20.32 & - & - & - \\
UK13 & S{\'e}rsic & 15.03 & 0.96 & 1.37 & 0.72 \\

\\

& PSF & 21.94 & - & - & - \\
UK14 & S{\'e}rsic & 16.71 & 0.99 & 0.89 & 0.64\\

\\

& Inner S{\'e}rsic & 18.33 & 3.87 & 0.31 & 0.91\\
UK15 & Outer S{\'e}rsic & 15.38 & 1.00 & 2.13 & 0.89 \\
\\

& PSF & 21.09 & - & - & - \\
UK19 & S{\'e}rsic & 16.31 & 1.80 & 0.97 & 0.73 \\

\\

& PSF & 20.13 & - & - & - \\
UK22 & S{\'e}rsic & 14.89 & 1.38 & 1.35 & 0.88\\
\\

& PSF & 23.36 & - & - & - & - \\
UK23 & Inner S{\'e}rsic & 18.18 & 0.26 & 0.54 & 0.67\\
& Outer S{\'e}rsic & 15.58 & 1.00 & 3.20 & 0.62\\
\\

& PSF & 24.13 & - & - & - \\
UK24 & S{\'e}rsic & 17.04 & 1.66 & 2.12 & 0.30 & \\
\\

& Inner S{\'e}rsic & 16.65 & 1.37 & 0.42 & 0.64 \\
UK26 & Outer S{\'e}rsic & 15.98 & 1.00 & 0.65 & 0.77 \\
\\

& PSF & 23.01 & - & - & - \\
UK27 & S{\'e}rsic & 17.29 & 0.98 & 1.96 & 0.76 &  \\

\\

& PSF & 20.66 & - & - & - \\
UK28 & S{\'e}rsic & 15.60 & 0.93 & 1.60 & 0.82 \\

\\

& PSF & 23.37 & - & - & - \\
UK29 & S{\'e}rsic & 17.63 & 0.66 & 1.64 & 0.48 \\

\\

& PSF & 22.53 & - & - & - \\
UK32 & S{\'e}rsic & 18.34 & 0.81 & 0.84 & 0.52 \\

\\

UK33 & S{\'e}rsic & 15.24 & 1.89 & 1.14 & 0.46 \\

\\

& PSF & 24.31 & - & - & - \\
UK34 & S{\'e}rsic & 16.05 & 1.40 & 2.69 & 0.40 & Unmodeled Spiral Arms\\

\\

& PSF & 22.62 & - & - & - \\
UK43 & S{\'e}rsic & 14.93 & 0.79 & 3.09 & 0.72 \\

\\

UK46 & S{\'e}rsic & 15.51 & 1.37 & 0.75 & 0.76\\

\\

UK47 & S{\'e}rsic & 16.56 & 1.15 & 0.92 & 0.89\\
\\

& PSF & 21.50 & - & - & - \\
UK49 & S{\'e}rsic & 14.80 & 1.65 & 1.82 & 0.55 \\

\\

& PSF & 21.60 & - & - & - \\
UK50 & Inner S{\'e}rsic & 16.07 & 1.51 & 1.39 & 0.34 \\
& Outer S{\'e}rsic & 14.70 & 1.00 & 2.54 & 0.46\\
\\

& PSF & 19.65 & - & - & - \\
UK51 & S{\'e}rsic & 15.00 & 0.92 & 1.03 & 0.45 \\
\\

& PSF & 20.01 & - & - & - \\
UK52 & S{\'e}rsic & 14.34 & 1.29 & 1.64 & 0.56 \\
\\

& PSF & 19.10 & - & - & - \\
UK55 & S{\'e}rsic & 14.47 & 1.18 & 1.24 & 0.77 & Unmodeled Spiral Arms \\
\\

& PSF & 23.50 & - & - & - \\
UK59 & Inner S{\'e}rsic & 16.79 & 1.03 & 0.85 & 0.57 \\
& Outer S{\'e}rsic & 16.26 & 1.00 & 2.64 & 0.47\\
\\

& PSF & 22.55 & - & - & - \\
UK60 & S{\'e}rsic & 15.35 & 0.47 & 1.36 & 0.55 \\

\\

& PSF & 21.71 & - & - & - \\
UK63 & S{\'e}rsic & 16.52 & 1.58 & 1.33 & 0.56 \\

\\

UK64 & S{\'e}rsic & 16.67 & 1.59 & 0.89 & 0.82 \\

\\
& PSF & 21.48 & - & - & - \\
UK67 & Inner S{\'e}rsic & 15.74 & 1.47 & 0.87 & 0.90 & Unmodeled Spiral Arms \\
& Outer S{\'e}rsic & 12.92 & 1.00 & 4.93 & 0.82 \\
\\

& PSF & 19.82 & - & - & - \\
UK72 & S{\'e}rsic & 14.58 & 3.41 & 1.81 & 0.62 \\

\\

& PSF & 22.65 & - & - & - \\
UK74 & Inner S{\'e}rsic & 18.28 & 0.93 & 0.47 & 0.28 \\
& Outer S{\'e}rsic & 16.11 & 1.00 & 1.24 & 0.95 \\
\\

& Inner S{\'e}rsic & 17.85 & 1.69 & 0.70 & 0.44 \\
UK77 & Outer S{\'e}rsic & 14.94 & 1.00 & 2.31 & 0.72 & Unmodeled Spiral Arms \\
\\

& PSF & 22.29 & - & - & - \\
UK78 & S{\'e}rsic & 14.79 & 0.60 & 1.04 & 0.55 \\
\\

UK82 & S{\'e}rsic & 16.12 & 1.07 & 3.00 & 0.38 \\

\enddata
\tablecomments{Column 1: Galaxy Identification Number. Column 2: Components in GALFIT model. Column 3: Total apparent magnitude in the UKIRT Y band. Column 4: Best fitting S{\'e}rsic index. Column 5: Best fitting effective radius, converted to kpc. Column 6: Best fitting axis ratio (b/a). Column 7: Any additional component included in the best-fit model. 
}
\end{deluxetable*}

\bibliography{refs}

\begin{thebibliography}{}
\expandafter\ifx\csname natexlab\endcsname\relax\def\natexlab#1{#1}\fi

\bibitem[{{Ba{\~n}ados} {et~al.}(2018){Ba{\~n}ados}, {Venemans},
  {Mazzucchelli}, {Farina}, {Walter}, {Wang}, {Decarli}, {Stern}, {Fan},
  {Davies}, {Hennawi}, {Simcoe}, {Turner}, {Rix}, {Yang}, {Kelson}, {Rudie}, \&
  {Winters}}]{banados2018}
{Ba{\~n}ados}, E., {Venemans}, B.~P., {Mazzucchelli}, C., {et~al.} 2018, \nat,
  553, 473

\bibitem[{{Baldassare} {et~al.}(2020){Baldassare}, {Geha}, \&
  {Greene}}]{baldassare2020}
{Baldassare}, V.~F., {Geha}, M., \& {Greene}, J. 2020, \apj, 896, 10

\bibitem[{{Baldry} {et~al.}(2012){Baldry}, {Driver}, {Loveday}, {Taylor},
  {Kelvin}, {Liske}, {Norberg}, {Robotham}, {Brough}, {Hopkins}, {Bamford},
  {Peacock}, {Bland-Hawthorn}, {Conselice}, {Croom}, {Jones}, {Parkinson},
  {Popescu}, {Prescott}, {Sharp}, \& {Tuffs}}]{gama2012}
{Baldry}, I.~K., {Driver}, S.~P., {Loveday}, J., {et~al.} 2012, \mnras, 421,
  621

\bibitem[{{Baldwin} {et~al.}(1981){Baldwin}, {Phillips}, \&
  {Terlevich}}]{baldwinetal1981}
{Baldwin}, J.~A., {Phillips}, M.~M., \& {Terlevich}, R. 1981, \pasp, 93, 5

\bibitem[{{Begelman} \& {Rees}(1978)}]{begelman1978}
{Begelman}, M.~C., \& {Rees}, M.~J. 1978, \mnras, 185, 847

\bibitem[{{Blanton} {et~al.}(2011){Blanton}, {Kazin}, {Muna}, {Weaver}, \&
  {Price-Whelan}}]{blanton2011}
{Blanton}, M.~R., {Kazin}, E., {Muna}, D., {Weaver}, B.~A., \& {Price-Whelan},
  A. 2011, \aj, 142, 31

\bibitem[{{Bruzual} \& {Charlot}(2003)}]{bruzualcharlot2003}
{Bruzual}, G., \& {Charlot}, S. 2003, \mnras, 344, 1000

\bibitem[{{Burke} {et~al.}(2023){Burke}, {Shen}, {Liu}, {Natarajan}, {Caplar},
  {Bellovary}, \& {Wang}}]{burke2023}
{Burke}, C.~J., {Shen}, Y., {Liu}, X., {et~al.} 2023, \mnras, 518, 1880

\bibitem[{{Carlsten} {et~al.}(2021){Carlsten}, {Greene}, {Beaton}, \&
  {Greco}}]{carlsten2021}
{Carlsten}, S.~G., {Greene}, J.~E., {Beaton}, R.~L., \& {Greco}, J.~P. 2021,
  arXiv e-prints, arXiv:2105.03440

\bibitem[{{de Vaucouleurs}(1948)}]{devauc}
{de Vaucouleurs}, G. 1948, Annales d'Astrophysique, 11, 247

\bibitem[{{den Brok} {et~al.}(2014){den Brok}, {Peletier}, {Seth}, {Balcells},
  {Dominguez}, {Graham}, {Carter}, {Erwin}, {Ferguson}, {Goudfrooij},
  {Guzm{\'a}n}, {Hoyos}, {Jogee}, {Lucey}, {Phillipps}, {Puzia}, {Valentijn},
  {Verdoes Kleijn}, \& {Weinzirl}}]{denBrokComa}
{den Brok}, M., {Peletier}, R.~F., {Seth}, A., {et~al.} 2014, \mnras, 445, 2385

\bibitem[{{Evans} {et~al.}(2010){Evans}, {Primini}, {Glotfelty}, {Anderson},
  {Bonaventura}, {Chen}, {Davis}, {Doe}, {Evans}, {Fabbiano}, {Galle}, {Gibbs},
  {Grier}, {Hain}, {Hall}, {Harbo}, {He}, {Houck}, {Karovska}, {Kashyap},
  {Lauer}, {McCollough}, {McDowell}, {Miller}, {Mitschang}, {Morgan},
  {Mossman}, {Nichols}, {Nowak}, {Plummer}, {Refsdal}, {Rots}, {Siemiginowska},
  {Sundheim}, {Tibbetts}, {Van Stone}, {Winkelman}, \& {Zografou}}]{evans2010}
{Evans}, I.~N., {Primini}, F.~A., {Glotfelty}, K.~J., {et~al.} 2010, \apjs,
  189, 37

\bibitem[{{Gebhardt} {et~al.}(2001){Gebhardt}, {Lauer}, {Kormendy}, {Pinkney},
  {Bower}, {Green}, {Gull}, {Hutchings}, {Kaiser}, {Nelson}, {Richstone}, \&
  {Weistrop}}]{gebhardt2001}
{Gebhardt}, K., {Lauer}, T.~R., {Kormendy}, J., {et~al.} 2001, \aj, 122, 2469

\bibitem[{Georgiev {et~al.}(2009)Georgiev, Hilker, Puzia, Goudfrooij, \&
  Baumgardt}]{georgiev2009}
Georgiev, I.~Y., Hilker, M., Puzia, T.~H., Goudfrooij, P., \& Baumgardt, H.
  2009, Monthly Notices of the Royal Astronomical Society, 396, 1075

\bibitem[{{Ghez} {et~al.}(2008){Ghez}, {Salim}, {Weinberg}, {Lu}, {Do}, {Dunn},
  {Matthews}, {Morris}, {Yelda}, {Becklin}, {Kremenek}, {Milosavljevic}, \&
  {Naiman}}]{ghez2008}
{Ghez}, A.~M., {Salim}, S., {Weinberg}, N.~N., {et~al.} 2008, \apj, 689, 1044

\bibitem[{{Graham} \& {Guzm{\'a}n}(2003)}]{coma2003}
{Graham}, A.~W., \& {Guzm{\'a}n}, R. 2003, \aj, 125, 2936

\bibitem[{Greene(2012)}]{Greene}
Greene, J.~E. 2012, Nature Communications, 3, 1304 EP , review Article

\bibitem[{{Greene} \& {Ho}(2007)}]{greeneho2007}
{Greene}, J.~E., \& {Ho}, L.~C. 2007, \apj, 670, 92

\bibitem[{Greene {et~al.}(2020)Greene, Strader, \& Ho}]{greene2020}
Greene, J.~E., Strader, J., \& Ho, L.~C. 2020, Annual Review of Astronomy and
  Astrophysics, 58, 257

\bibitem[{{Habouzit} {et~al.}(2017){Habouzit}, {Volonteri}, \&
  {Dubois}}]{habouzit2016}
{Habouzit}, M., {Volonteri}, M., \& {Dubois}, Y. 2017, \mnras, 468, 3935

\bibitem[{{Inayoshi} {et~al.}(2020){Inayoshi}, {Visbal}, \&
  {Haiman}}]{inayoshi2020}
{Inayoshi}, K., {Visbal}, E., \& {Haiman}, Z. 2020, \araa, 58, 27

\bibitem[{{Jiang} {et~al.}(2011){Jiang}, {Greene}, {Ho}, {Xiao}, \&
  {Barth}}]{jiang}
{Jiang}, Y.-F., {Greene}, J.~E., {Ho}, L.~C., {Xiao}, T., \& {Barth}, A.~J.
  2011, \apj, 742, 68

\bibitem[{{Karachentsev} {et~al.}(2013){Karachentsev}, {Makarov}, \&
  {Kaisina}}]{karachentsev}
{Karachentsev}, I.~D., {Makarov}, D.~I., \& {Kaisina}, E.~I. 2013, \aj, 145,
  101

\bibitem[{{Kelvin} {et~al.}(2012){Kelvin}, {Driver}, {Robotham}, {Hill},
  {Alpaslan}, {Baldry}, {Bamford}, {Bland-Hawthorn}, {Brough}, {Graham},
  {H{\"a}ussler}, {Hopkins}, {Liske}, {Loveday}, {Norberg}, {Phillipps},
  {Popescu}, {Prescott}, {Taylor}, \& {Tuffs}}]{kelvin2012}
{Kelvin}, L.~S., {Driver}, S.~P., {Robotham}, A. S.~G., {et~al.} 2012, \mnras,
  421, 1007

\bibitem[{{Kimbrell} {et~al.}(2021){Kimbrell}, {Reines}, {Schutte}, {Greene},
  \& {Geha}}]{kimbrell2021}
{Kimbrell}, S.~J., {Reines}, A.~E., {Schutte}, Z., {Greene}, J.~E., \& {Geha},
  M. 2021, \apj, 911, 134

\bibitem[{{Kormendy}(2015)}]{kormendy2015dwarf}
{Kormendy}, J. 2015, in Lessons from the Local Group: A Conference in honor of
  David Block and Bruce Elmegreen, 323

\bibitem[{{Kormendy} {et~al.}(2011){Kormendy}, {Bender}, \&
  {Cornell}}]{kormendy2011}
{Kormendy}, J., {Bender}, R., \& {Cornell}, M.~E. 2011, \nat, 469, 374

\bibitem[{{Kormendy} \& {Ho}(2013)}]{kormendy}
{Kormendy}, J., \& {Ho}, L.~C. 2013, \araa, 51, 511

\bibitem[{{Kormendy} \& {Richstone}(1995)}]{kormendy1995}
{Kormendy}, J., \& {Richstone}, D. 1995, \araa, 33, 581

\bibitem[{{Lawrence} {et~al.}(2007){Lawrence}, {Warren}, {Almaini}, {Edge},
  {Hambly}, {Jameson}, {Lucas}, {Casali}, {Adamson}, {Dye}, {Emerson},
  {Foucaud}, {Hewett}, {Hirst}, {Hodgkin}, {Irwin}, {Lodieu}, {McMahon},
  {Simpson}, {Smail}, {Mortlock}, \& {Folger}}]{Lawrence}
{Lawrence}, A., {Warren}, S.~J., {Almaini}, O., {et~al.} 2007, \mnras, 379,
  1599

\bibitem[{{Loeb} \& {Rasio}(1994)}]{loeb1994}
{Loeb}, A., \& {Rasio}, F.~A. 1994, \apj, 432, 52

\bibitem[{{Madau} \& {Rees}(2001)}]{madau2001}
{Madau}, P., \& {Rees}, M.~J. 2001, \apjl, 551, L27

\bibitem[{Massey(1951)}]{massey1951}
Massey, F.~J. 1951, Journal of the American Statistical Association, 46, 68

\bibitem[{{McConnachie}(2012)}]{mcconnachie}
{McConnachie}, A.~W. 2012, \aj, 144, 4

\bibitem[{{Miller} {et~al.}(2015){Miller}, {Gallo}, {Greene}, {Kelly}, {Treu},
  {Woo}, \& {Baldassare}}]{miller2015}
{Miller}, B.~P., {Gallo}, E., {Greene}, J.~E., {et~al.} 2015, \apj, 799, 98

\bibitem[{{Mortlock} {et~al.}(2011){Mortlock}, {Warren}, {Venemans}, {Patel},
  {Hewett}, {McMahon}, {Simpson}, {Theuns}, {Gonz{\'a}les-Solares}, {Adamson},
  {Dye}, {Hambly}, {Hirst}, {Irwin}, {Kuiper}, {Lawrence}, \&
  {R{\"o}ttgering}}]{Mortlock}
{Mortlock}, D.~J., {Warren}, S.~J., {Venemans}, B.~P., {et~al.} 2011, \nat,
  474, 616

\bibitem[{{Neumayer} {et~al.}(2020){Neumayer}, {Seth}, \&
  {B{\"o}ker}}]{neumayer}
{Neumayer}, N., {Seth}, A., \& {B{\"o}ker}, T. 2020, \aapr, 28, 4

\bibitem[{{Oh} {et~al.}(2017){Oh}, {Greene}, \& {Lackner}}]{oh2017}
{Oh}, S., {Greene}, J.~E., \& {Lackner}, C.~N. 2017, \apj, 836, 115

\bibitem[{{Peng} {et~al.}(2010){Peng}, {Ho}, {Impey}, \& {Rix}}]{peng}
{Peng}, C.~Y., {Ho}, L.~C., {Impey}, C.~D., \& {Rix}, H.-W. 2010, \aj, 139,
  2097

\bibitem[{Reines(2022)}]{reines2022}
Reines, A.~E. 2022, Nature Astronomy, 6, 26

\bibitem[{{Reines} {et~al.}(2013){Reines}, {Greene}, \& {Geha}}]{reines}
{Reines}, A.~E., {Greene}, J.~E., \& {Geha}, M. 2013, \apj, 775, 116

\bibitem[{{Ricarte} \& {Natarajan}(2018)}]{ricarte2018}
{Ricarte}, A., \& {Natarajan}, P. 2018, \mnras, 481, 3278

\bibitem[{{S{\'a}nchez-Janssen} {et~al.}(2019){S{\'a}nchez-Janssen},
  {C{\^o}t{\'e}}, {Ferrarese}, {Peng}, {Roediger}, {Blakeslee}, {Emsellem},
  {Puzia}, {Spengler}, {Taylor}, {{\'A}lamo-Mart{\'\i}nez}, {Boselli},
  {Cantiello}, {Cuillandre}, {Duc}, {Durrell}, {Gwyn}, {MacArthur},
  {Lan{\c{c}}on}, {Lim}, {Liu}, {Mei}, {Miller}, {Mu{\~n}oz}, {Mihos},
  {Paudel}, {Powalka}, \& {Toloba}}]{sanchezjanssen2019}
{S{\'a}nchez-Janssen}, R., {C{\^o}t{\'e}}, P., {Ferrarese}, L., {et~al.} 2019,
  \apj, 878, 18

\bibitem[{{Satyapal} {et~al.}(2009){Satyapal}, {B{\"o}ker}, {Mcalpine},
  {Gliozzi}, {Abel}, \& {Heckman}}]{satyapal2009}
{Satyapal}, S., {B{\"o}ker}, T., {Mcalpine}, W., {et~al.} 2009, \apj, 704, 439

\bibitem[{{Schutte} {et~al.}(2019){Schutte}, {Reines}, \& {Greene}}]{schutte}
{Schutte}, Z., {Reines}, A.~E., \& {Greene}, J.~E. 2019, \apj, 887, 245

\bibitem[{{S{\'e}rsic}(1963)}]{sersic}
{S{\'e}rsic}, J.~L. 1963, Boletin de la Asociacion Argentina de Astronomia La
  Plata Argentina, 6, 41

\bibitem[{{She} {et~al.}(2017){She}, {Ho}, \& {Feng}}]{she2017chandra}
{She}, R., {Ho}, L.~C., \& {Feng}, H. 2017, \apj, 842, 131

\bibitem[{{Vanden Berk} {et~al.}(2001){Vanden Berk}, {Richards}, {Bauer},
  {Strauss}, {Schneider}, {Heckman}, {York}, {Hall}, {Fan}, {Knapp},
  {Anderson}, {Annis}, {Bahcall}, {Bernardi}, {Briggs}, {Brinkmann}, {Brunner},
  {Burles}, {Carey}, {Castander}, {Connolly}, {Crocker}, {Csabai}, {Doi},
  {Finkbeiner}, {Friedman}, {Frieman}, {Fukugita}, {Gunn}, {Hennessy},
  {Ivezi{\'c}}, {Kent}, {Kunszt}, {Lamb}, {Leger}, {Long}, {Loveday}, {Lupton},
  {Meiksin}, {Merelli}, {Munn}, {Newberg}, {Newcomb}, {Nichol}, {Owen}, {Pier},
  {Pope}, {Rockosi}, {Schlegel}, {Siegmund}, {Smee}, {Snir}, {Stoughton},
  {Stubbs}, {SubbaRao}, {Szalay}, {Szokoly}, {Tremonti}, {Uomoto}, {Waddell},
  {Yanny}, \& {Zheng}}]{vandenberk2001}
{Vanden Berk}, D.~E., {Richards}, G.~T., {Bauer}, A., {et~al.} 2001, \aj, 122,
  549

\bibitem[{Vito {et~al.}(2017)Vito, Brandt, Yang, Gilli, Luo, Vignali, Xue,
  Comastri, Koekemoer, Lehmer, Liu, Paolillo, Ranalli, Schneider, Shemmer,
  Volonteri, \& Wang}]{vito}
Vito, F., Brandt, W.~N., Yang, G., {et~al.} 2017, Monthly Notices of the Royal
  Astronomical Society, 473, 2378

\bibitem[{Volonteri(2010)}]{volonteri}
Volonteri, M. 2010, The Astronomy and Astrophysics Review, 18, 279

\bibitem[{{Volonteri} {et~al.}(2021){Volonteri}, {Habouzit}, \&
  {Colpi}}]{volonteri2021}
{Volonteri}, M., {Habouzit}, M., \& {Colpi}, M. 2021, Nature Reviews Physics,
  3, 732

\bibitem[{{York} {et~al.}(2000){York}, {Adelman}, {Anderson}, {Anderson},
  {Annis}, {Bahcall}, {Bakken}, {Barkhouser}, {Bastian}, {Berman}, {Boroski},
  {Bracker}, {Briegel}, {Briggs}, {Brinkmann}, {Brunner}, {Burles}, {Carey},
  {Carr}, {Castander}, {Chen}, {Colestock}, {Connolly}, {Crocker}, {Csabai},
  {Czarapata}, {Davis}, {Doi}, {Dombeck}, {Eisenstein}, {Ellman}, {Elms},
  {Evans}, {Fan}, {Federwitz}, {Fiscelli}, {Friedman}, {Frieman}, {Fukugita},
  {Gillespie}, {Gunn}, {Gurbani}, {de Haas}, {Haldeman}, {Harris}, {Hayes},
  {Heckman}, {Hennessy}, {Hindsley}, {Holm}, {Holmgren}, {Huang}, {Hull},
  {Husby}, {Ichikawa}, {Ichikawa}, {Ivezi{\'c}}, {Kent}, {Kim}, {Kinney},
  {Klaene}, {Kleinman}, {Kleinman}, {Knapp}, {Korienek}, {Kron}, {Kunszt},
  {Lamb}, {Lee}, {Leger}, {Limmongkol}, {Lindenmeyer}, {Long}, {Loomis},
  {Loveday}, {Lucinio}, {Lupton}, {MacKinnon}, {Mannery}, {Mantsch}, {Margon},
  {McGehee}, {McKay}, {Meiksin}, {Merelli}, {Monet}, {Munn}, {Narayanan},
  {Nash}, {Neilsen}, {Neswold}, {Newberg}, {Nichol}, {Nicinski}, {Nonino},
  {Okada}, {Okamura}, {Ostriker}, {Owen}, {Pauls}, {Peoples}, {Peterson},
  {Petravick}, {Pier}, {Pope}, {Pordes}, {Prosapio}, {Rechenmacher}, {Quinn},
  {Richards}, {Richmond}, {Rivetta}, {Rockosi}, {Ruthmansdorfer}, {Sandford},
  {Schlegel}, {Schneider}, {Sekiguchi}, {Sergey}, {Shimasaku}, {Siegmund},
  {Smee}, {Smith}, {Snedden}, {Stone}, {Stoughton}, {Strauss}, {Stubbs},
  {SubbaRao}, {Szalay}, {Szapudi}, {Szokoly}, {Thakar}, {Tremonti}, {Tucker},
  {Uomoto}, {Vanden Berk}, {Vogeley}, {Waddell}, {Wang}, {Watanabe},
  {Weinberg}, {Yanny}, {Yasuda}, \& {SDSS Collaboration}}]{york2000}
{York}, D.~G., {Adelman}, J., {Anderson}, John~E., J., {et~al.} 2000, \aj, 120,
  1579

\end{thebibliography}

\end{document}